\documentclass[apj]{emulateapj}
\usepackage{color}

\newcommand{\eg}{e.g., }
\newcommand{\ie}{i.e., }
\newcommand{\Msun}{M_{\odot}}
\newcommand{\Rsun}{R_{\odot}}
\newcommand{\kms}{km~s$^{-1}$}

\newcommand{\Cofs}{$^{56}$Co}
\newcommand{\Nifs}{$^{56}$Ni}
\newcommand{\Mms}{M_{\rm MS}}
\newcommand{\Mej}{M_{\rm ej}}
\newcommand{\KE}{E_{\rm K}}
\newcommand{\Mni}{M({\rm ^{56}Ni})}
\def\gsim{\mathrel{\rlap{\lower 4pt \hbox{\hskip 1pt $\sim$}}\raise 1pt
\hbox {$>$}}}
\def\lsim{\mathrel{\rlap{\lower 4pt \hbox{\hskip 1pt $\sim$}}\raise 1pt
\hbox {$<$}}}
\newcommand{\vph}{v_{\rm ph}}

\def\ion#1#2{{\rm #1}~{\sc #2}}

\shorttitle{Type Ib Supernova 2008D}
\shortauthors{Tanaka et al.}
 
\begin{document}

\title{Type I\lowercase{b} Supernova 2008D associated with the Luminous X-ray Transient 080109:
An Energetic Explosion of a Massive Helium Star
}
\author{
Masaomi Tanaka\altaffilmark{1,2},
Nozomu Tominaga\altaffilmark{3,1},
Ken'ichi Nomoto\altaffilmark{2,1},
S. Valenti\altaffilmark{4},
D.K. Sahu\altaffilmark{5},
T. Minezaki\altaffilmark{6},
Y. Yoshii\altaffilmark{6,7},
M. Yoshida\altaffilmark{8},
G.C. Anupama\altaffilmark{5},
S. Benetti\altaffilmark{9},
G. Chincarini\altaffilmark{10,11},
M. Della Valle\altaffilmark{12,13},
P. A. Mazzali\altaffilmark{14,9}, and
E. Pian\altaffilmark{15}
}

\altaffiltext{1}{Department of Astronomy, School of Science, University of Tokyo, 7-3-1 Hongo, Bunkyo-ku, Tokyo 113-0033, Japan; mtanaka@astron.s.u-tokyo.ac.jp}
\altaffiltext{2}{Institute for the Physics and Mathematics of the Universe, University of Tokyo, Kashiwanoha 5-1-5, Kashiwa, Chiba 277-8568, Japan}
\altaffiltext{3}{Division of Optical and Infrared Astronomy, National Astronomical Observatory of Japan, 2-21-1 Osawa, Mitaka, Tokyo 181-8588, Japan; nozomu.tominaga@nao.ac.jp}

\altaffiltext{4}{Astrophysics Research Centre, School of Maths and Physics, Queen's University, Belfast, BT7 1NN, Northern Ireland, UK}
\altaffiltext{5}{Indian Institute of Astrophysics, II Block Koramangala, Bangalore 560034, India}
\altaffiltext{6}{Institute of Astronomy, School of Science, University of Tokyo, 2-21-1 Osawa, Mitaka, Tokyo 181-0015, Japan}
\altaffiltext{7}{Research Center for the Early Universe, School of Science, University of Tokyo, 7-3-1 Hongo, Bunkyo-ku, Tokyo 113-003, Japan}
\altaffiltext{8}{Okayama Astrophysical Observatory, National Astronomical Observatory of Japan}
\altaffiltext{9}{Istituto Naz. di Astrofisica-Oss. Astron., vicolo dell'Osservatorio, 5, 35122 Padova, Italy}
\altaffiltext{10}{Universit degli Studi di Milano Bicocca, Dipartimento di Fisica, Piazza della Scienze 3, 20126 Milano, Italy}
\altaffiltext{11}{INAF, Osservatorio Astronomico di Brera, via E. Bianchi 46, 23807 Merate (LC), Italy}
\altaffiltext{12}{Capodimonte Astronomical Observatory,  Salita Moiariello 16, I-80131, INAF- Napoli, Italy}
\altaffiltext{13}{European Southern Observatory, Karl-Schwarzschild-Strasse 2, D-85748, Garching, Germany}
\altaffiltext{14}{Max-Planck Institut f\"ur Astrophysik, Karl-Schwarzschild-Strasse 2 D-85748 Garching bei M\"unchen, Germany}
\altaffiltext{15}{Istituto Naz. di Astrofisica-Oss. Astron., Via Tiepolo, 11, 34131 Triste, Italy}

\begin{abstract}
We present a theoretical model for supernova (SN) 2008D associated 
with the luminous X-ray transient 080109.
The bolometric light curve and optical spectra of the SN are modelled based on 
the progenitor models and the explosion models obtained from 
hydrodynamic/nucleosynthetic calculations.
We find that SN 2008D is a more energetic explosion than normal 
core-collapse supernovae, with an ejecta mass of $\Mej = 5.3 \pm 1.0\ \Msun$
and a kinetic energy of $\KE = 6.0 \pm 2.5 \times 10^{51}$ erg.
The progenitor star of the SN has a $6-8 \Msun$ He core 
with essentially no H envelope ($< 5 \times 10^{-4}\ \Msun$)
prior to the explosion.
The main-sequence mass of the progenitor
is estimated to be $\Mms =20-25\ \Msun$, 
with additional systematic 
uncertainties due to convection, mass loss, rotation, and binary effects.
These properties are intermediate between those of 
normal SNe and hypernovae associated with gamma-ray bursts.
The mass of the central remnant is estimated as $1.6-1.8 \Msun$, which 
is near the boundary between neutron star and black hole formation.
\end{abstract}

\keywords{supernovae: general --- supernovae: individual (SN~2008D) --- 
nuclear reactions, nucleosynthesis, abundances --- radiative transfer}

\section{Introduction}
\label{sec:intro}

A luminous X-ray transient was discovered in NGC 2770
in the {\it Swift} XRT data taken 
on 2008 January 9 for the observation of SN 2007uy in the same galaxy
(Berger \& Soderberg 2008; Kong \& Maccarone 2008).
The X-ray emission of 
the transient reached a peak $\sim 65$ seconds, lasting $\sim 600$ seconds,
after the observation started (Page et al. 2008).
The X-ray spectrum is soft, and no $\gamma$-ray counterpart was detected by
the {\it Swift} BAT (Page et al. 2008).

Given the small total X-ray energy and the soft X-ray emission,
Soderberg et al. (2008) and Chevalier \& Fransson (2008) 
interpreted the X-ray transient as a supernova (SN) shock breakout.
On the other hand, Xu et al. (2008), Li (2008) and Mazzali et al. (2008)
considered this transient as the least energetic end of 
gamma-ray bursts (GRBs) and X-ray flashes (XRFs).

The optical counterpart was discovered at the position of 
the X-ray transient (Deng \& Zhu 2008; Valenti et al. 2008a),
confirming the presence of a SN, named SN 2008D (Li \& Filippenko 2008).
To study this event in detail, intensive follow-up observations were carried 
out over a wide wavelength range including X-rays 
(Soderberg et al. 2008; Modjaz et al. 2008b), 
optical/NIR (Soderberg et al. 2008; Malesani et al. 2009; 
Modjaz et al. 2008b; Mazzali et al. 2008), and radio (Soderberg et al. 2008).

\begin{deluxetable*}{lrrrrrrrrrr} 
%\rotate
\tablewidth{0pt}
\tablecaption{Explosion Models}
\tablehead{
Model & 
$\Mms$ \tablenotemark{a} & 
$M_{\alpha}$ \tablenotemark{b} & 
$R_*$ \tablenotemark{c} & 
$M_{\rm cut}$ \tablenotemark{d} & 
$\Mej$ \tablenotemark{e} &
$\KE$ \tablenotemark{f} &
$M($\Nifs$)$ \tablenotemark{g} &
$v_{\rm He}$ \tablenotemark{h} &
$v_{\rm Ni}$ \tablenotemark{i} &
$M_{0.1c}$ \tablenotemark{j} 
}
\startdata
HE4          & $\approx$ 15 & 4  & 3.5  & 1.3 & 2.7 & 1.1 & 0.07 
& $<$3500 & 7900 &  $< 3.0 \times 10^{-5}$  \\
HE6          & $\approx$ 20 & 6  & 2.2  & 1.6 & 4.4 & 3.7 & 0.065  
& 6700    & 7000 &  0.007   \\
HE8          & $\approx$ 25 & 8  & 1.3  & 1.8 & 6.2 & 8.4 & 0.07 
& 10500   & 9000 &  0.04   \\ 
HE10         & $\approx$ 30 & 10 & 1.2  & 2.3 & 7.7 & 13.0 & 0.07 
& 12500   & 10600&  0.09   \\ 
HE16         & $\approx$ 40 & 16 & 0.74  & 3.6 & 12.4& 26.5 & 0.07 
& 17500   & 14000&  0.24   \\ 
\hline
\multicolumn{2}{l}{Soderberg et al. (2008)} & & $\sim 1$ \tablenotemark{k} 
&  & 3-5     &  2-4    & 0.05  &   & &  \\ 
Mazzali et al. (2008) & $\sim 30$  & & 
&  & $\sim$7 & $\sim$6 & 0.09  &   & & 0.03
\enddata
\tablenotetext{a}{Main-sequence mass ($\Msun$) estimated 
from the approximate formula obtained by Sugimoto \& Nomoto (1980)} 
\tablenotetext{b}{The mass of the He star ($\Msun$)} 
\tablenotetext{c}{Progenitor radius prior to the explosion ($\Rsun$)} 
\tablenotetext{d}{Mass cut ($\Msun$)} 
\tablenotetext{e}{The mass of the SN ejecta ($\Msun$)}
\tablenotetext{f}{The kinetic energy of the SN ejecta ($10^{51}$ erg)}
\tablenotetext{g}{The mass of ejected \Nifs\ ($\Msun$)}
\tablenotetext{h}{Velocity at the bottom of the He layer (\kms)}
\tablenotetext{i}{Velocity at the outer boundary of \Nifs\ distribution (\kms)}
\tablenotetext{j}{The ejecta mass at $v > 0.1c$ ($\Msun$)}
\tablenotetext{k}{Estimated from the photospheric radius and temperature
of the early part of the LC ($t \lsim 4$ days)
using the formulae by Waxman et al. (2007).
This is consistent with the estimate by Modjaz et al. (2008b) while 
Chevalier \& Fransson (2008) derived $\sim 9 \Rsun$ using 
$\Mej$ and $\KE$ estimated by Soderberg et al. (2008) 
and the formulae by Chevalier (1992).}
\label{tab:models}
\end{deluxetable*}

SN 2008D showed a broad-line optical spectrum 
at early epochs ($t \lsim 10$ days, hereafter $t$ denotes time 
after the transient, 2008 Jan 9.56 UT; Soderberg et al. 2008).
However, the spectrum changed to that of normal Type Ib SN, \ie 
SN with He absorption lines and without H lines (Modjaz et al. 2008c).
To date, the SN associated with GRBs or XRFs are all Type Ic,
\ie SNe without H and He absorption.

Soderberg et al. (2008) and Mazzali et al. (2008) 
estimated the ejected mass and the kinetic energy of SN 2008D.
Using analytic formulae, Soderberg et al. (2008) suggested that this SN has
the ejecta mass $\Mej = 3 - 5 \Msun$ 
and the kinetic energy of the ejecta $\KE \sim 2 - 4 \times 10^{51}$ erg.
Mazzali et al. (2008) did model calculations and suggested that this event is 
intermediate between normal SNe and GRB-associated SNe (or hypernovae), 
with $\Mej \sim 7 \Msun$ and $\KE \sim 6 \times 10^{51}$ erg.
In their model calculations, hydrodynamic/nucleosynthetic models are not 
used, and thus, their estimate of the core mass prior to the explosion
and the progenitor main-sequence mass is less direct.

In this paper, we present detailed theoretical study of 
emission from SN 2008D.
The bolometric light curve (LC) and optical spectra are modelled 
based on the progenitor models and the explosion models obtained from 
hydrodynamic/nucleosynthetic calculations.
In \S \ref{sec:models}, we show the progenitor and explosion models.
The optical LC and spectra are modelled in \S \ref{sec:LC} 
and \S \ref{sec:spec}, respectively.
We discuss the nature of SN 2008D in \S \ref{sec:discussion}
and finally give conclusions in \S \ref{sec:conclusions}.
Throughout this paper, we adopt 31 Mpc ($\mu = 32.46$ mag) 
for the distance to SN 2008D (Modjaz et al. 2008b; Mazzali et al. 2008)
and $E(B-V)=0.65$ mag for the total reddening (Mazzali et al. 2008).

\section{Models}
\label{sec:models}

To understand the nature of SN 2008D and its progenitor star,
(1) we first construct the exploding He star models
with various masses by performing hydrodynamic/nucleosynthetic calculations 
for the presupernova He star models.
(2) The important parameters of the SN, such as the mass ($\Mej$)
and kinetic energy of the ejecta ($\KE$),
are estimated by modelling the bolometric LC (\S \ref{sec:LC}) and 
the optical spectra (\S \ref{sec:spec}).
Then, (3) the He star mass ($M_{\alpha}$) can be estimated from the 
best set of $\Mej$ and $\KE$.
Finally, (4) the main-sequence mass ($\Mms$) of the progenitor star
can be estimated by the evolution models, 
which predict the $\Mms$-$M_{\alpha}$ relation.

In this section, we construct hydrodynamic models using an evolutionary model.
The progenitor model and hydrodynamic/nucleosynthetic calculations
are described in \S \ref{sec:progenitor} and \S \ref{sec:hydro}, respectively.
The hydrodynamic models are tested in \S \ref{sec:LC} and \S \ref{sec:spec}.

\subsection{Progenitor Models}
\label{sec:progenitor}

In the strategy described above,
presupernova He star models are required as input for
the hydrodynamic calculations.
The Wolf-Rayet star models with stellar winds tend to form the 
He stars whose masses are larger than that inferred for
SN 2008D (\eg Soderberg et al. 2008).
To study the properties of the ejecta and the progenitor star
without specifying the mass loss mechanism (stellar winds 
in Wolf-Rayet star or Roche lobe overflow in close binary, \ie
possible binary progenitor scenario, \eg Wellstein \& Langer 1999),
we adopt the He star evolution models with various masses.
We use five He star models with 
the masses of $4$, $6$, $8$, $10$ and $16 \Msun$ 
(Nomoto \& Hashimoto 1988; Nomoto et al. 1997; Nakamura et al. 2001b),
where the mass loss is not taken into account.
These models are called HE4, HE6, HE8, HE10 and HE16, respectively.
The corresponding main-sequence masses of these models are 
$\approx 15$, $20$, $25$, $30$ and $40 \Msun$, 
respectively (Table \ref{tab:models}),
which is estimated from the approximate formula 
of the $\Mms$-$M_{\alpha}$ relation obtained by 
Sugimoto \& Nomoto (1980; Eq. 4.1).

The difference in the density structure of the He core is negligible 
among different stellar evolutionary calculations 
if the He core mass is the same.
As a result, the observable quantities (\ie LC and spectra)
after the hydrodynamic and radiative transfer calculations
are not affected by the variety of the evolutionary models.
Thus, the estimate of $\Mej$ and $\KE$ does not depend on 
the evolutionary models.

\begin{figure}
\begin{center}
%\epsscale{1.0}
%\plotone{f1.eps}
\includegraphics[scale=0.75]{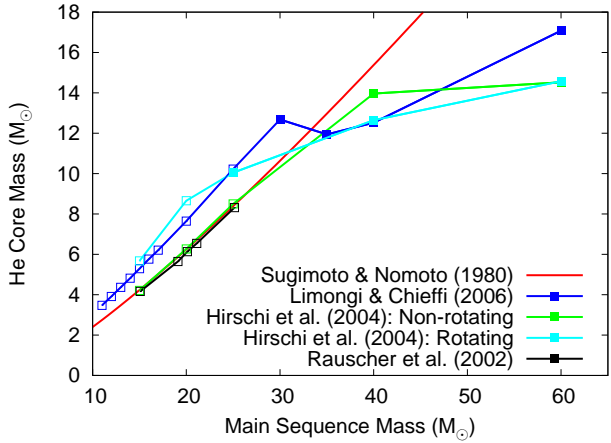}
\caption{Relation between the main-sequence mass 
and the He core mass in the formula derived by Sugimoto \& Nomoto 
(1980; red, used in Nomoto \& Hashimoto 1988).
It is compared with the relation between the main-sequence mass 
and the He core mass at the presupernova stage in the models by
Limongi \& Chieffi (2006, blue), Hirschi et al. (2004, green and cyan for 
non-rotating and rotating models, respectively), 
and Rauscher et al. (2002, black).
The models shown in open squares have H envelope while 
the models shown in filled squares have a bare He core.
\label{fig:ms_core}}
\end{center}
\end{figure}

However, the $\Mms$-$M_{\alpha}$ relation
depends on the several evolutionary processes that are subject to 
uncertainties, \eg 
the convective overshooting, wind mass loss, shear mixing and
meridional circulation in rotating stars.
The different assumptions adopted in different stellar calculation
codes may affect the $\Mms$-$M_{\alpha}$ relation
as summarized in Figure \ref{fig:ms_core}.
The red line shows the formula derived by Sugimoto \& Nomoto (1980).
The other lines show
the relations obtained by
the models including mass loss (Limongi \& Chieffi 2006,
blue; Hirschi et al. 2004, green; Rauscher et al. 2002, black), and
rotation ($v_{ZAMS}=300$ \kms; Hirschi et al. 2004, cyan).
The models shown in open squares have a H envelope
prior to the explosion while
the models shown in the filled squares have a bare He core.

These models assume the solar abundance for the initial abundance.
The mass loss causes the smaller $M_{\alpha}$ for $\Mms> 30 \Msun$ 
in the models by Limongi \& Chieffi (2006) and Hirschi et al. (2004).
For the lower metallicity models, the mass loss
rate is lower, so that the $\Mms$-$M_{\alpha}$ relation would be closer to
that of Sugimoto \& Nomoto (1980).
For the stars with $\Mms \lsim 30 \Msun$ or $M_{\alpha} \lsim 10 \Msun$,
the gradient in the plot is similar among the models.

At the presupernova stage, the He stars consist of the Fe core, 
Si-rich layer, O-rich layer, and He-rich layer.
The mass of the Fe-core is $\sim 1.4 - 1.6 \Msun$, depending on the model.
The mass of the O-rich layer is sensitive to the progenitor mass, while 
the mass of the He-rich layer is $\sim 2 \Msun$
irrespectively of the He star mass.
Note that the mass of He-rich layer can be as large as $\sim 3 \Msun$ 
depending on the evolutionary models (\eg Limongi \& Chieffi 2006) and 
can also be smaller than $2 \Msun$ prior to the explosion by mass loss.

The mass fraction of O in the O-rich layer is $\sim 0.8$.
Other abundant elements in this layer are Ne, Mg, and C,
with mass fractions of order 0.1.
These are almost irrespective of the evolutionary models.
The He mass fraction in the He-rich layer is $\sim 0.9$.
The second most abundant element in this layer is C, with a mass fraction 
of $\sim 0.03$, but this is rather uncertain (\S \ref{sec:LC}).
Oxygen is also produced in the He-rich layer, 
but the mass fraction of O is only $\sim 0.01$.

\begin{figure}
\begin{center}
%\epsscale{1.0}
%\plotone{f1.eps}
\includegraphics[scale=1.1]{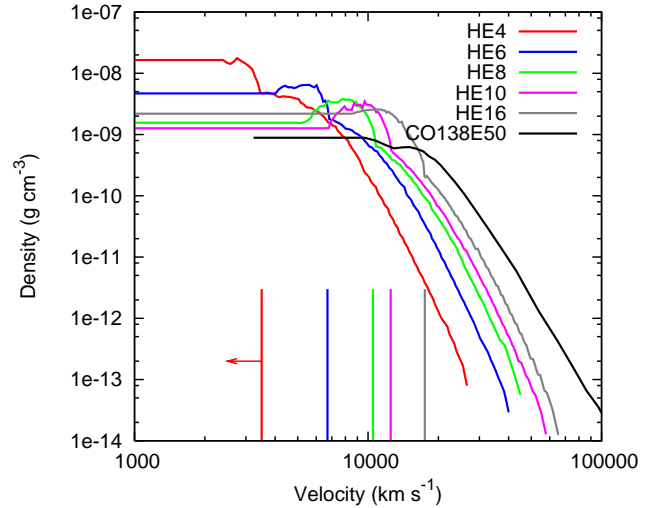}
\caption{
Density profile of the explosion models at one day after the explosion.
Red (HE4), blue (HE6), green (HE8), magenta (HE10), and gray (HE16)
lines show the models for SN 2008D (see Table \ref{tab:models}). 
The black line shows the C+O star explosion model 
used for SN 1998bw in Nakamura et al. (2001a). 
The vertical lines show the velocity at the bottom of the He layer
in each model (with the same colors).
\label{fig:dens}}
\end{center}
\end{figure}

\subsection{Hydrodynamics \& Nucleosynthesis}
\label{sec:hydro}

The hydrodynamics of the SN explosion and explosive nucleosynthesis 
are calculated for the five progenitor models.
The hydrodynamic calculations are performed 
by a spherical Lagrangian hydrodynamic code 
with the piecewise parabolic method (PPM; Colella \& Woodward 1984).
The code includes nuclear energy production from the $\alpha$ network.
The equation of state includes gas, radiation, e$^-$-e$^+$ pairs,
Coulomb interaction between ions and electrons and phase transition 
(Nomoto 1982; Nomoto \& Hashimoto 1988).
The explosion is initiated by increasing the temperatures
at a few meshes below the mass cut (see below), \ie a thermal bomb.

The SN ejecta become homologous at $\sim 1000$ s after the explosion.
After the hydrodynamic calculations, nucleosynthesis is calculated 
for each model as a post-processing (Hix \& Thielemann 1996, 1999).
The reaction network includes 280 isotopes up to $^{79}$Br.
The results of the nucleosynthesis depends on the progenitor mass
and the kinetic energy of the explosion.
The kinetic energies in the five models are determined to explain the 
observed LC (\S \ref{sec:LC}).

The explosion models are summarized in Table \ref{tab:models}.
The mass cut ($M_{\rm cut}$) is defined after the nucleosynthesis
calculation to eject the optimal amount of \Nifs\ to power the LC.
Figure \ref{fig:dens} shows the density structure of the explosion models
at one day after the initiation of the explosion. 
The ``bump'' in the density profile is caused by the reverse shock
generated at the boundary of the C+O/He layers.

The vertical lines in Figure \ref{fig:dens} show the velocity at the bottom 
of the He layer after the expansion of SN ejecta become homologous
($v_{\rm He}$, Table \ref{tab:models}).
Since strong mixing is expected in less massive stars 
($M_{\alpha} \lsim 4 \Msun$, Hachisu et al. 1991), the value of $v_{\rm He}$ 
in model HE4 is the upper limit of the inner velocity of the He-rich layer.
If the mass of the He layer prior to the explosion is 
larger (smaller) than $\sim 2 \Msun$ as in our model set, 
$v_{\rm He}$ can be lower (higher).

\section{Bolometric Light Curve}
\label{sec:LC}

\begin{figure*}
\begin{center}
\begin{tabular}{cc}
\includegraphics[scale=1.1]{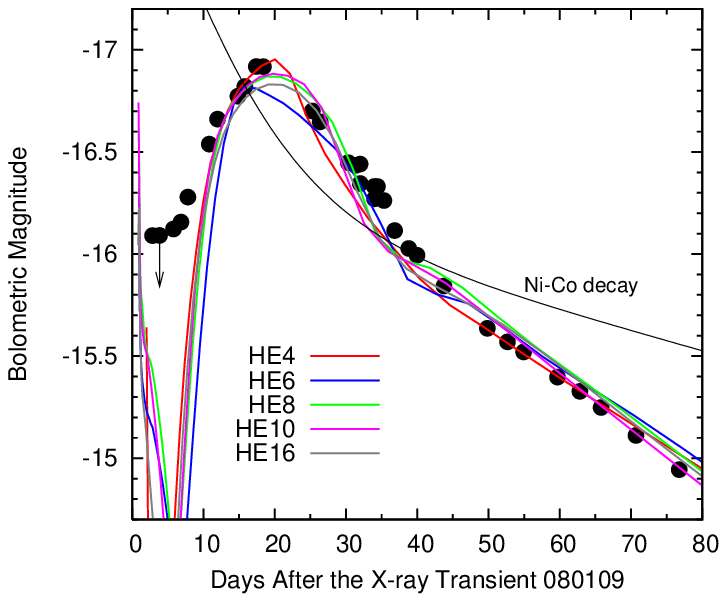}&
\includegraphics[scale=1.1]{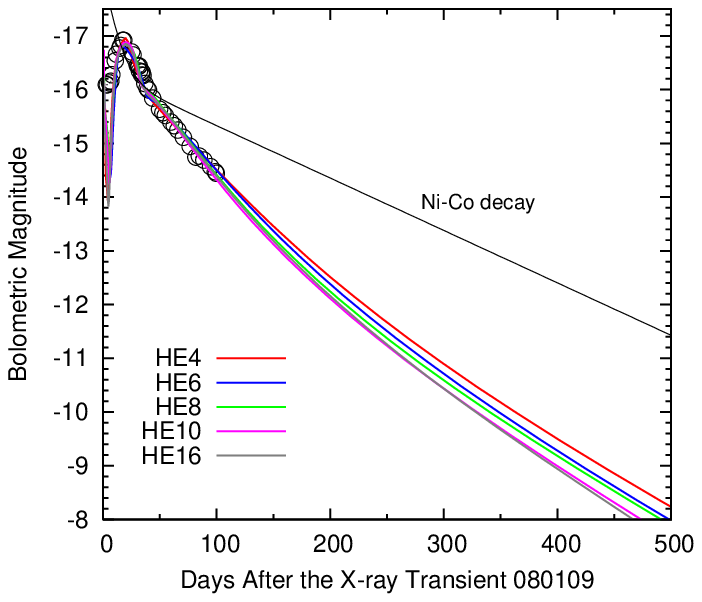}
\end{tabular}
\caption{
Pseudo-bolometric ($UBVRIJHK$) LC of SN 2008D 
(Minezaki et al. in preparation)
compared with the results of LC calculations with the models HE4 (red), 
HE6(blue), HE8 (green), HE10 (magenta) and HE16 (gray).
The pseudo-bolometric LC is shown in filled (left) 
and open (right) circles.
The thin black line shows the decay energy from \Nifs\ and \Cofs\ 
[$M$(\Nifs) $=$ 0.07 $\Msun$].
The bolometric magnitude at $t \sim 4$ days after the X-ray 
transient is brighter by $\sim 0.25$ mag 
than that shown by other papers (Soderberg et al. 2008;
Malesani et al. 2009; Modjaz et al. 2008b; Mazzali et al. 2008),
which is shown by the thin arrow in the left panel (see Appendix \ref{app:LC}).
\label{fig:LC}}
\end{center}
\end{figure*}

The pseudo-bolometric ($UBVRIJHK$) LC was constructed by Minezaki et al. 
(in preparation, see Appendix \ref{app:LC}) compiling optical data taken by the MAGNUM telescope 
(Yoshii 2002; Yoshii, Kobayashi \& Minezaki 2003), 
the Himalayan Chandra Telescope, and {\it Swift} UVOT 
($U$-band, Soderberg et al. 2008),
and also NIR data taken by the MAGNUM telescope.
The first part of the LC ($t \lsim 4$ days) seems to be related 
to the X-ray transient or the subsequent tail 
(Soderberg et al. 2008; Chevalier \& Fransson 2008) while 
the later part ($t \gsim 4$ days) is the SN component, 
powered by the decay of \Nifs\ and \Cofs.

The first part of the LC depends on the progenitor radius and 
radiation-hydrodynamics at outer layers, as well as 
$\Mej$ and $\KE$.
To determine the global properties of the SN ejecta, 
we focus on the second, principal part, which depends on
$\Mej$, $\KE$ and the amount of ejected \Nifs\ mass [$\Mni$].
The progenitor radius is discussed in \S \ref{sec:comp}.

The LCs are calculated for the five explosion
models presented in \S \ref{sec:models} (see Table \ref{tab:models}).
Our LTE, time-dependent radiative transfer code (Iwamoto et al. 2000) 
solves the Saha equation to obtain the ionization structure.
Using the calculated electron density, the Rossland mean opacity is 
calculated approximately by the empirical relation to the electron scattering 
opacity derived from the TOPS database (Magee et al. 1995, Deng et al. 2005).
For the initial temperature structure of the SN ejecta, we use results of 
adiabatic hydrodynamic calculations at one day after the explosion.
The hydrodynamics and the radiative transfer are not coupled.

Asphericity of the ejecta of SN 2008D is suggested by the emission line 
profile in the spectrum at $t = 109$ days (Modjaz et al. 2008b).
To include the possible effect of aspherical explosion, we modify the 
distribution of \Nifs\ from that derived from nucleosynthetic calculation.
In hydrodynamic/nucleosynthetic calculations of aspherical explosion, 
more \Nifs\ is mixed to the surface in the more aspherical cases 
(see \eg Maeda et al. 2006, Tominaga 2009).
A constant mass fraction of \Nifs\ is assumed below 
$v_{\rm Ni}$, the outer boundary of \Nifs\ distribution in velocity.
The value of $v_{\rm Ni}$ is determined so as to explain the rising part 
of the LC.
The estimated $v_{\rm Ni}$ is listed in Table \ref{tab:models}.
The resultant mass fraction of \Nifs\ is from 0.03 (HE4) to 0.01 (HE16).

Figure \ref{fig:LC} shows the calculated LCs compared with the observed LC. 
The model LCs of HE8, HE10, and HE16 reproduce the observed 
LC around the peak very well.
The LCs of HE4 and HE6 tend to be narrower than the observations. 
At a later phase, the five LCs are all in good agreement with the observations.
The steep decline in the calculated LCs at $t \lsim 4$ days 
could be a relic of the 
shock-heated envelope, and radiation-hydrodynamics calculations 
are required to study this part.

HE4 and HE6 need some enhancement of C in the He layer 
to reproduce the observed LC near the peak more nicely.
The C-abundance in the He layer is poorly known because of the uncertainties 
involved in the C-production by convective 3 $\alpha$-reaction in progenitor
models and those in the Rayleigh-Taylor instability at the He/C+O interface 
during explosions, which tends to be stronger for lower mass He stars 
(Hachisu et al. 1991).  In view of these uncertainties, we include HE4 and HE6 
in the further spectral analysis, rather than excluding them from 
the possible models.

The timescale around the peak depends on both $\Mej$ and $\KE$
as $\propto \kappa^{1/2} \Mej^{3/4} \KE^{-1/4}$, 
where $\kappa$ is the optical opacity (Arnett 1982).
Thus, for each model, a kinetic energy can be specified
so as to reproduce the observed timescale.
The derived set of ejecta parameters are ($\Mej/\Msun$, $\KE/10^{51}$ erg) =
(2.7, 1.1), (4.4, 3.7), (6.2, 8.4), (7.7, 13.0) and (12.4, 26.5)
for the case of HE4, HE6, HE8, HE10 and HE16, respectively.
The ejected \Nifs\ mass is $\sim 0.07 \Msun$ in all models.

The model with $\Mej= 3-5 \Msun$ and $\KE = 2-4 \times 10^{51}$ erg 
suggested 
by Soderberg et al. (2008) is close to our HE6 model, while the model of 
Mazzali et al. (2008), with 
$\Mej= 7 \Msun$ and $\KE = 6 \times 10^{51}$ erg is close to our HE8 model.
Model HE4 has the canonical explosion energy of core-collapse SNe 
(\ie $\sim 10^{51}$ erg)
while HE10 and HE16 have the explosion energy of hypernovae ($> 10^{52}$ erg).

In all five models, the late evolution of the LC ($t>200$ days)
is not very different, with a decline rate of $\sim 0.015$ mag day$^{-1}$ 
(right panel of Fig. \ref{fig:LC}).
This decline is faster than the \Cofs\ decay rate (0.01 mag day$^{-1}$, 
thin black line in Fig. \ref{fig:LC}) because some $\gamma$-rays escape 
without depositing energy in the SN ejecta at such late epochs.
These models predict that the optical magnitude of SN 2008D 
is $\sim -10.5$ mag 
(observed magnitude $\sim 23.8$ with no bolometric correction) 
in 2008 October, 
\ie $\sim 300$ days after the explosion, when the SN can be observed again,
and $\sim -9.6$ mag ($\sim 24.7$ mag) at one year after the explosion
(if dust does not form in the ejecta).

\begin{figure}
\begin{center}
\includegraphics[scale=1.1]{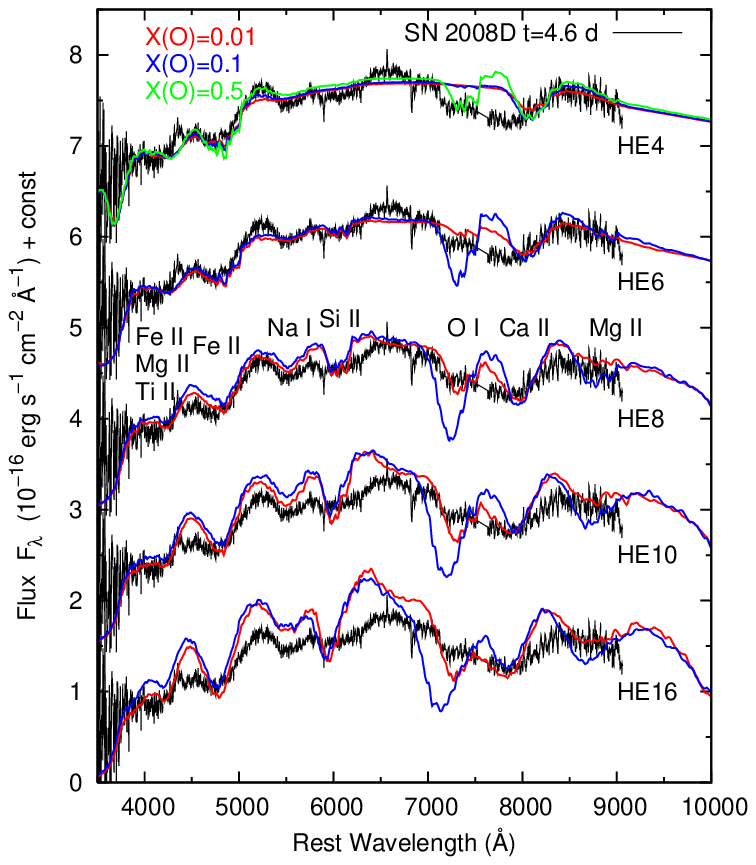}
\caption{
Spectrum of SN 2008D at $t=4.6$ days from the X-ray transient
(black line, Mazzali et al. 2008) 
compared with synthetic spectra (color lines).
The spectra are shifted by 6.0, 4.5, 3.0, 1.5, 0.0 from top to bottom.
The model spectra are reddened with $E(B-V)=0.65$ mag.
From top to bottom, the synthetic spectra calculated with HE4, HE6, HE8,
HE10, and HE16 are shown.
The red, blue and green lines show the synthetic spectra with oxygen 
mass fraction $X$(O) = 0.01, 0.1, and 0.5, respectively.
Since the synthetic spectra with $X$(O) = 0.1 for more massive models
than HE4 already show too strong \ion{O}{i} line, the spectra with $X$(O) = 0.5
are not shown for these models.
\label{fig:t4}}
\end{center}
\end{figure}

\section{Optical Spectra}
\label{sec:spec}

In this section, the five models are tested against the observed spectra.
Optical spectra have been shown by Soderberg et al. (2008), 
Malesani et al. (2009), Modjaz et al. (2008b) and Mazzali et al. (2008).
We use the data set presented by Mazzali et al. (2008).
The spectral sequence can be divided into three parts.
At the earliest epochs ($t \lsim 4$ days), 
the spectra are almost featureless
\footnote{Two absorption features are identified
around 4000 \AA\ in the spectra at $t \sim 2$ days 
($t=1.77$ days, Malesani et al. 2009; $t=1.84$ days, Modjaz et al. 2008b),
while they are not seen in the spectra at $t=1.54$ and $2.49$ days 
presented by Mazzali et al. (2008).
These absorptions might be due to more highly-ionized ions, such as 
\ion{C}{iii}, \ion{N}{iii}. and \ion{O}{iii}
(Modjaz et al. 2008b; Quimby et al. 2007).
We have investigated these lines 
by the Monte Carlo spectrum synthesis code, but 
we don't find a large contribution of these ions
because ionizations by the photospheric radiation only is not enough 
for the strong contributions of such ions,
as noted by Modjaz et al. (2008b).}.
This is probably the result of shock heating (the first part of the LC).
At $4 \lsim t \lsim 10$ days, the spectra show broad-line features.
Around and after maximum ($t \gsim 10$ days), the spectrum shows strong He 
features as in Type Ib SNe.
The velocity of the He lines is $\sim 9000$ - $10,000$ \kms\ 
(\S \ref{sec:vel}).
We present spectral modelling at the SN dominated phase, \ie $t \gsim 4$ days.

For spectral modelling, we use the one-dimensional
Monte Carlo spectrum synthesis code (Mazzali \& Lucy 1993).
The code assumes a spherically symmetric, sharply defined photosphere.
Electron and line scattering are taken into account.
For line scattering, the effect of line branching is included
(Lucy 1999; Mazzali 2000).
The ionization structure is calculated with modified nebular approximation
as in Mazzali \& Lucy (1993, see also Abbott \& Lucy 1985).
Although it is known that the non-thermal excitation is important for the He 
lines (Lucy 1991), non-thermal processes are not included in our analysis.
Thus, we do not aim to obtain a good fit of the He lines.

To determine the temperature structure, many photon packets are first 
traced above the photosphere with an assumed temperature structure.
The Monte Carlo ray tracing gives the flux at each mesh and 
the temperature structure is then updated using the flux.
This procedure is repeated until the temperature converges.
Finally a model spectrum is obtained using a formal integral (Lucy 1999).

The input parameters of the code are emergent luminosity ($L$), 
the position of the photosphere in velocity 
(photospheric velocity, $\vph$), and element abundances (mass fractions)
above the photosphere (\ie in the SN atmosphere).
Note that $L$ and $\vph$ do not depend much on the model parameters
such as $\Mej$ and $\KE$.
They are constrained by the absolute flux of the spectrum and the 
line velocities, respectively 
(and also by the relation of $L \propto \vph^2 t^2 T_{\rm eff}^4$,
where $T_{\rm eff}$ is the effective temperature of the spectrum).

With the estimated luminosity and photospheric velocity, 
mass fractions of elements are optimized.
For simplicity, homogeneous abundances are assumed above the 
photosphere without using the results of nucleosynthetic calculations.
We compare the derived abundances with those by nucleosynthetic calculations
for the progenitor models.
The goodness of the fit is judged by eyes
because of the complex dependences of the parameters
and the difficulty in obtaining the perfect fit of the overall spectrum.

\subsection{Broad-Line Spectrum: At $t = 4.6$ Days}
\label{sec:broad}

We first perform model calculations for the spectrum at $t=4.6$ days
(Fig. \ref{fig:t4}).
The spectrum shows broad-line features.

\subsubsection{Intermediate Mass Model HE8}

We use model HE8, the middle of our model sequence, as a fiducial case.
A good agreement with the observed spectrum is obtained with
$\vph = 18,500$ \kms\ and log $L$ (erg s$^{-1}$) = 41.7.
Since this velocity is higher than the He line velocities observed 
later phases ($\sim 9000$-$10,000$ \kms), 
the photosphere at this epoch is expected to be located in the He-rich layer.

Figure \ref{fig:t4} shows a comparison of the observed and synthetic spectrum.
The spectrum has P-Cygni profiles of \ion{O}{i}, \ion{Na}{i}, 
\ion{Ca}{ii}, \ion{Ti}{ii}, \ion{Cr}{ii}, and \ion{Fe}{ii} lines.
The line at 6000 \AA\ is identified as \ion{Si}{ii}.
The contribution of the high velocity H$\alpha$ is quite small, 
which is discussed in Appendix \ref{app:H}.
The spectrum at wavelengths bluer than 5500\AA\
is dominated by \ion{Ti}{ii}, \ion{Cr}{ii}, and \ion{Fe}{ii} lines.
Given the uncertainty in the metal abundances in outer layers, 
reflecting the uncertainty of the explosion mechanism or the degree of mixing,
these features can be fitted using the optimal value of the metal abundances.

In contrast to the heavy, synthesized elements, the oxygen abundance cannot 
be totally parameterized because the majority of oxygen is synthesized 
during the evolution of the progenitor star.
The red and blue line shows synthetic spectra with oxygen abundance
$X$(O)=0.01 and 0.1, respectively.
In these models, the abundance of He is $X$(He)$\sim$0.8 and 0.7, respectively.
The spectrum with $X$(O)=0.01 (red) gives a good match 
with the observed \ion{O}{i}$\lambda$7774 line around 7400\AA, while the 
\ion{O}{i} line in the model spectrum with $X$(O)=0.1 (blue) is too strong.
Since the oxygen abundance in the He-rich layer is of the order of $10^{-2}$
almost irrespective of evolutionary models, 
this is consistent with the fact that the 
photosphere is located in the He-rich layer.

In the observed spectrum, the \ion{O}{i} and \ion{Ca}{ii} IR triplet are 
blended at 7000 - 8500 \AA\ while they are separated in the synthetic spectra.
This is caused by the insufficient \ion{Ca}{ii} absorption in the model at 
the very high velocity layer with $v \sim 0.1c$.
The ejecta mass at $v> 0.1c$ in HE8 is $0.04 \Msun$ 
(Table \ref{tab:models}), 
which is consistent with that in the model presented by Mazzali et al. (2008).
Note that the mass at $v>0.1c$ is much smaller than in the model for 
SN 1998bw ($\sim 1.5 \Msun$, CO138E50 in Nakamura et al. 2001a, see also
Fig. \ref{fig:dens}).

\subsubsection{Massive Models HE10 and HE16}

Next, we use the more massive models.
For model HE10, the \ion{O}{i} line in the model with $X$(O)=0.01 (red) seems 
to be strong, but we can obtain a good fit with slightly smaller oxygen 
abundance.
For model HE16, the strength of the \ion{O}{i} line with $X$(O)=0.01 (red) is
similar to that in HE10.
In these massive models, the \ion{O}{i} feature is too strong in the model 
spectra with $X$(O)=0.1 (blue).
This is consistent with the fact that the photosphere ($v=18,500$ \kms) is 
located in the He-rich layer.

\subsubsection{Less Massive Models HE6 and HE4}

Finally, we use less massive models.
For the less massive model HE6, $X$(O)=0.01 gives a reasonable fit to 
the \ion{O}{i} line.
In contrast, $X$(O)=0.1 yields too strong a line.
This is a similar behavior to the more massive models, and 
implies that the photosphere is located in the He-rich layer.

For HE4, the synthetic spectra with $X$(O)=0.01 and 0.1 
do not give a strong enough \ion{O}{i} absorption (red and blue lines).
To explain the observed absorption, $X$(O)=0.5 is required (green line)
because of the low density at the outer layer of HE4 (Fig. \ref{fig:dens}).
This requires that the layer at $v= 18,500$ \kms\ should already be O-rich, 
which is clearly inconsistent with the observed He line velocity
($v \sim 9000$-$10,000$ \kms).
Therefore, HE4 is not likely to be a viable model for SN 2008D.

\begin{figure}
\begin{center}
\includegraphics[scale=1.1]{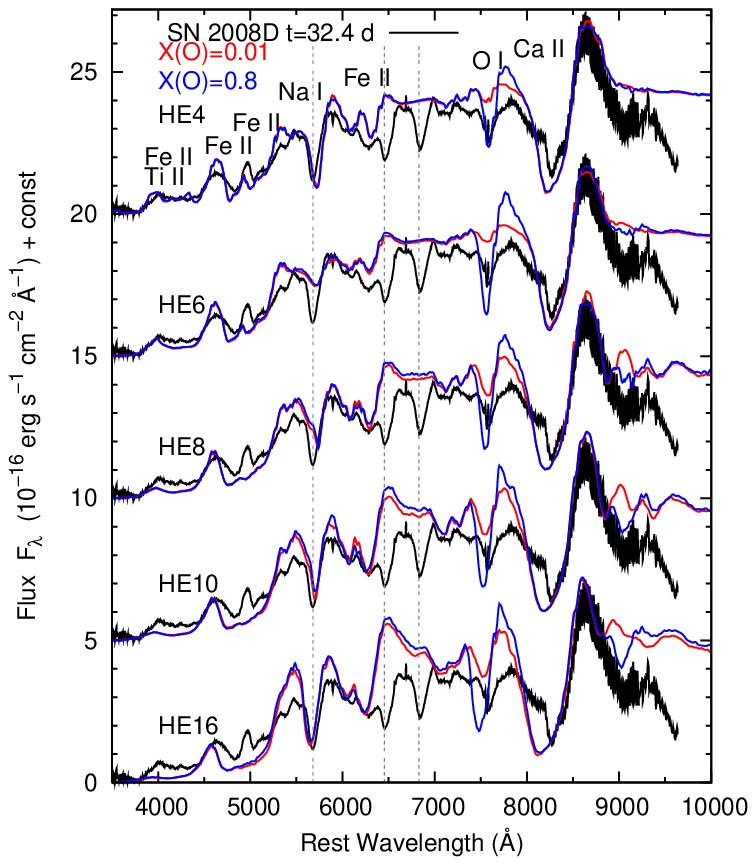}
\caption{
Spectrum of SN 2008D at $t=32.4$ days from the X-ray transient
(black line, Mazzali et al. 2008) compared with 
synthetic spectra (color lines).
The spectra are shifted by 20.0, 15.0, 10.0, 5.0, 0.0 from top to bottom.
The model spectra are reddened with $E(B-V)=0.65$ mag.
From top to bottom, the synthetic spectra calculated with HE4, HE6, HE8,
HE10, and HE16 are shown.
The red and blue lines show the synthetic spectra with an oxygen 
mass fraction of $X$(O) = 0.01 and 0.8, respectively.
Dashed, vertical 
lines show the position of the He lines (\ion{He}{i} 5876, 6678, 7065)
blueshifted with $v=10,000$ \kms.
Note that the He lines are not treated in the code.
\label{fig:t32}}
\end{center}
\end{figure}

\subsection{Type Ib Spectrum: At $t = 32.4$ Days }
\label{sec:Ib}

At $t=32.4$ days, the observed spectrum shows typical Type Ib features.
The overall features are fitted well with 
$\vph = 7500$\,\kms\ and log $L$ (erg s$^{-1}$) = 42.1 (Fig. \ref{fig:t32}).
The observed Fe lines at 4500 - 5000\,\AA\ are too narrow to be reproduced 
by the massive models (HE8, HE10 and HE16). This is caused by our crude 
assumption of a homogeneous abundance distribution, which is not appropriate  
for synthesized elements such as Fe in the inner layer.
The spectra might be improved using a stratified abundance distribution 
or non-spherical models (Tanaka et al. 2007).

\subsubsection{Intermediate Mass Model HE8}

The model spectrum with $X$(O)=0.01, as assumed for the spectrum at t=4.6 days,
is shown in red line.
The synthetic \ion{O}{i} line at 7500 \AA\ is 
slightly weaker than the observation.
A value $X$(O)=0.8 yields a reasonably strong \ion{O}{i} line (blue), which 
implies that the photosphere ($v =7500$ \kms) is not located in the He-rich 
layer, consistent with $v_{\rm He} = 10500$ \kms\
(velocity at the bottom of the He layer) of HE8.

\subsubsection{Massive Models HE10 and HE16}
\label{sec:t32massive}

The synthetic spectra calculated using HE10 and HE16 show 
the stronger \ion{O}{i} line than HE8.
The spectrum with $X$(O)=0.01 (red) gives a slightly weaker \ion{O}{i} line
than in the observation, while
the spectrum with $X$(O)=0.8 (blue) yields a sufficiently strong line.
Although the synthetic \ion{O}{i} line with $X$(O)=0.8 
is too strong especially at high velocity (\ie at bluer wavelength),
this is caused by the assumption of the homogeneous abundance distribution.
Thus, near the photosphere, a high mass fraction of O is preferred.
This is consistent with the high $v_{\rm He}$ of these models
($12,500$ and $17,500$ for HE10 and HE16, respectively).

However, the observed He line velocities ($v \sim 9000$-$10,000$ \kms)
suggest that the layer at $v \sim 10,000$ \kms\ is still He-rich.
This is inconsistent with the high $v_{\rm He}$ in HE10 and HE16,
requiring that the layers at $v \sim 10,000$ \kms\ be O-rich.

\subsubsection{Less Massive Models HE6 and HE4}

\begin{figure}
\begin{center}
\includegraphics[scale=1.]{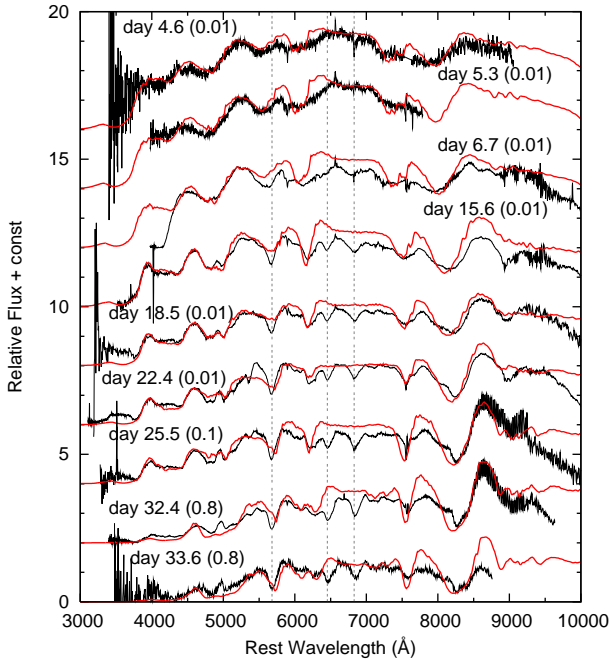}
\caption{
Spectral evolution of SN 2008D (black, Mazzali et al. 2008) compared with 
the sequence of the synthetic spectra computed with HE8 (red).
The epoch in the figure shows the days from the X-ray transient.
The values in the parenthesis shows the mass fraction of oxygen 
assumed in the calculation.
The spectra are shifted by 16.0, 14.0, 12.0, 10.0, 8.0, 6.0, 
4.0, 2.0, 0.0 from top to bottom.
The model spectra are reddened with $E(B-V)=0.65$ mag.
Dashed, vertical lines show the position of the He lines 
(\ion{He}{i} 5876, 6678, 7065)
blueshifted with $v=10,000$ \kms.
Note that the He lines are not treated in the code.
\label{fig:spec_seq}}
\end{center}
\end{figure}

The synthetic spectra using HE6 also have similar trend with those of HE8.
The spectrum with $X$(O)=0.8 (blue) 
gives a reasonable fit to the \ion{O}{i} line.
Although the low velocity at the bottom of the He layer in HE6 
($v_{\rm He}=6700$ \kms) suggests that the photosphere at this epoch 
($v=7500$ \kms) is still in the He layer, this small difference is within 
the uncertainty of $v_{\rm He}$ caused by the variation of the 
He layer mass depending on evolutionary models.

For HE4, the \ion{O}{i} absorption is reproduced with $X$(O)$=0.8$.
However, the very low $v_{\rm He}$ of HE4 ($<3500$ \kms) is  
not consistent with the fact that the spectrum model 
requires the O-dominated photosphere at $v=7500$ \kms.

\subsection{Velocity Evolution}
\label{sec:vel}

\begin{figure}
\begin{center}
\includegraphics[scale=1.1]{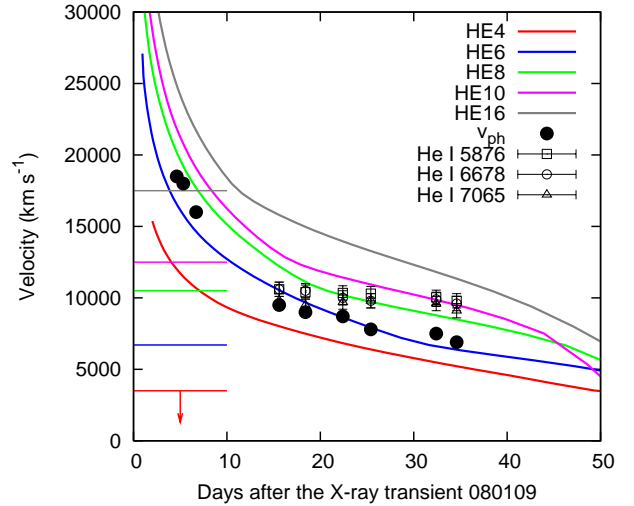}
\caption{
Time evolution of photospheric velocity calculated by the LC code 
(color lines) and that derived from the spectral modelling (filled circles).
Open black symbols show the line velocity of He lines.
The horizontal lines show $v_{\rm He}$ for each model (Table \ref{tab:models}).
Note that $v_{\rm He}$ can be varied depending on evolutionary models 
with different masses of the He layer.
\label{fig:vph}}
\end{center}
\end{figure}

Using the fiducial model HE8, we calculate the spectral evolution
(Fig. \ref{fig:spec_seq}).
The values in the parenthesis is the oxygen mass fraction adopted 
in the fitting.
We find that a higher oxygen mass fraction is preferred for the later
spectra.
Although a homogeneous mass fraction is assumed in the calculation,
the photospheric position seems to transit from the He-rich layer
to the O-rich layer around $t=25.5$ days.
Thus, the boundary between the He-rich and O-rich layers
is located near $v = 7800$ \kms.

Figure \ref{fig:vph} shows the photospheric velocities derived 
from the spectral modelling (filled black circles), which does not depend 
much on the model parameters.
In Figure \ref{fig:vph}, the photospheric velocities obtained from the 
synthetic LCs (\S \ref{sec:LC}) are also shown (solid lines).
The photospheric velocities for HE6 and HE8 are close to the values derived 
from spectral modelling.
However, it should be noted that the photospheric velocities obtained from 
the LC models are only approximate because the LC model assumes LTE 
and does not fully take into account the contribution of the line opacity.
Thus, uncertainty of a few thousand \kms\ is expected.
Nevertheless, the decreasing trend of the photospheric velocity
derived from the spectral modelling is reproduced by our LC calculations
because (1) we use the hydrodynamic models (Fig. \ref{fig:dens})
having a decreasing density structure toward the outer layers
and (2) we solve the ionization in the ejecta, 
and thus, the opacity is time-dependent.

In Figure \ref{fig:vph}, the Doppler velocities of three \ion{He}{i} lines 
measured at the absorption minimum (open symbols) are also shown.
Malesani et al. (2009) and Modjaz et al. (2008b) show the subsequent spectral 
evolution, and the He line velocity declines slowly to $v \sim 9000$ \kms.

The horizontal lines in Figure \ref{fig:vph} mark the velocity at the bottom 
of the He layer for the five models ($v_{\rm He}$, see Table \ref{tab:models}).
In HE10 and HE16, $v_{\rm He}$ is too high compared with the observed 
velocities (\S \ref{sec:t32massive}).
Also in HE8, it may be higher than the minimum of the observed 
He line velocity ($v \sim 9000$ \kms).
The lower $v_{\rm He}$ in HE4 and HE6 cannot be excluded 
from the observed line velocities.
But the spectral modelling shows that the layer at $v \sim 7500$ \kms\ 
is not He-rich (\S 4.2).
This is inconsistent with the very low $v_{\rm He}$ in HE4.
It must be cautioned that $v_{\rm He}$ is affected by the mass of the He layer
(\ie by the choice of evolutionary models).

\section{Discussion}
\label{sec:discussion}

\begin{figure}
\begin{center}
\includegraphics[scale=0.7]{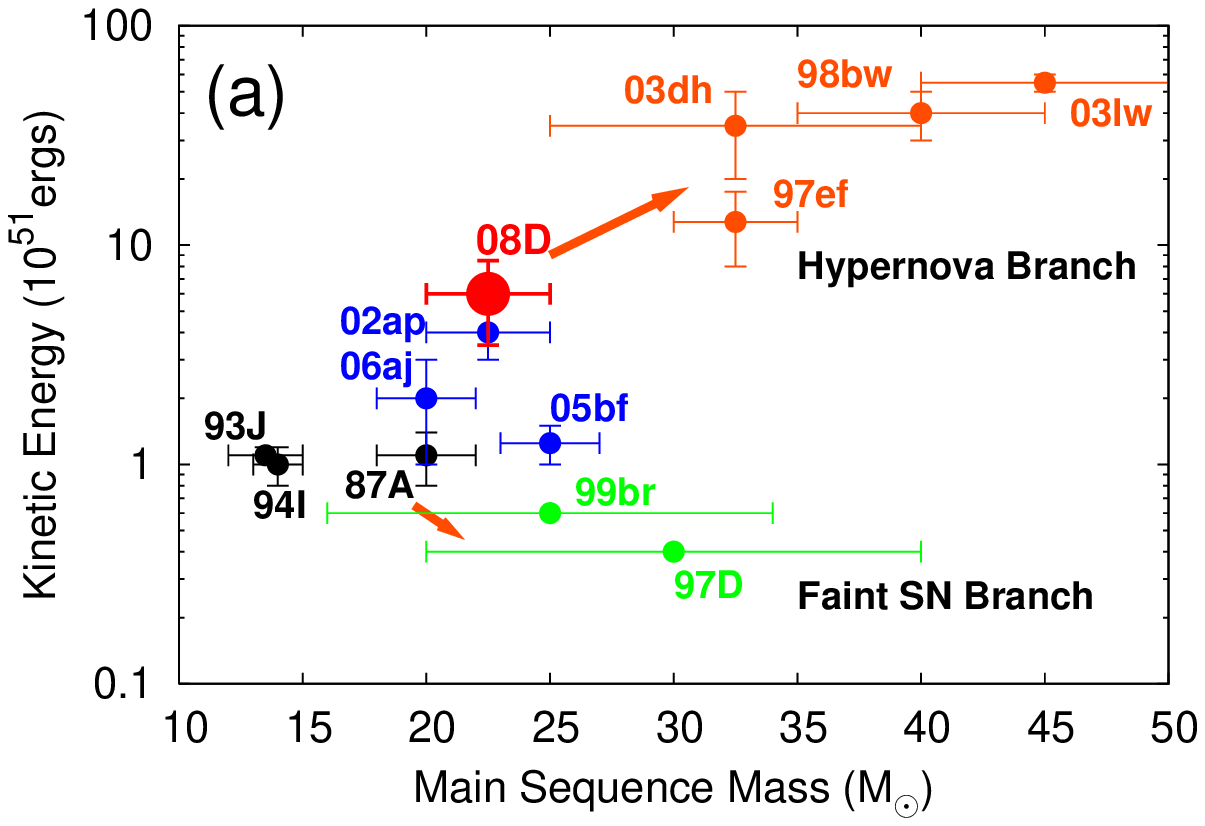}
\includegraphics[scale=0.7]{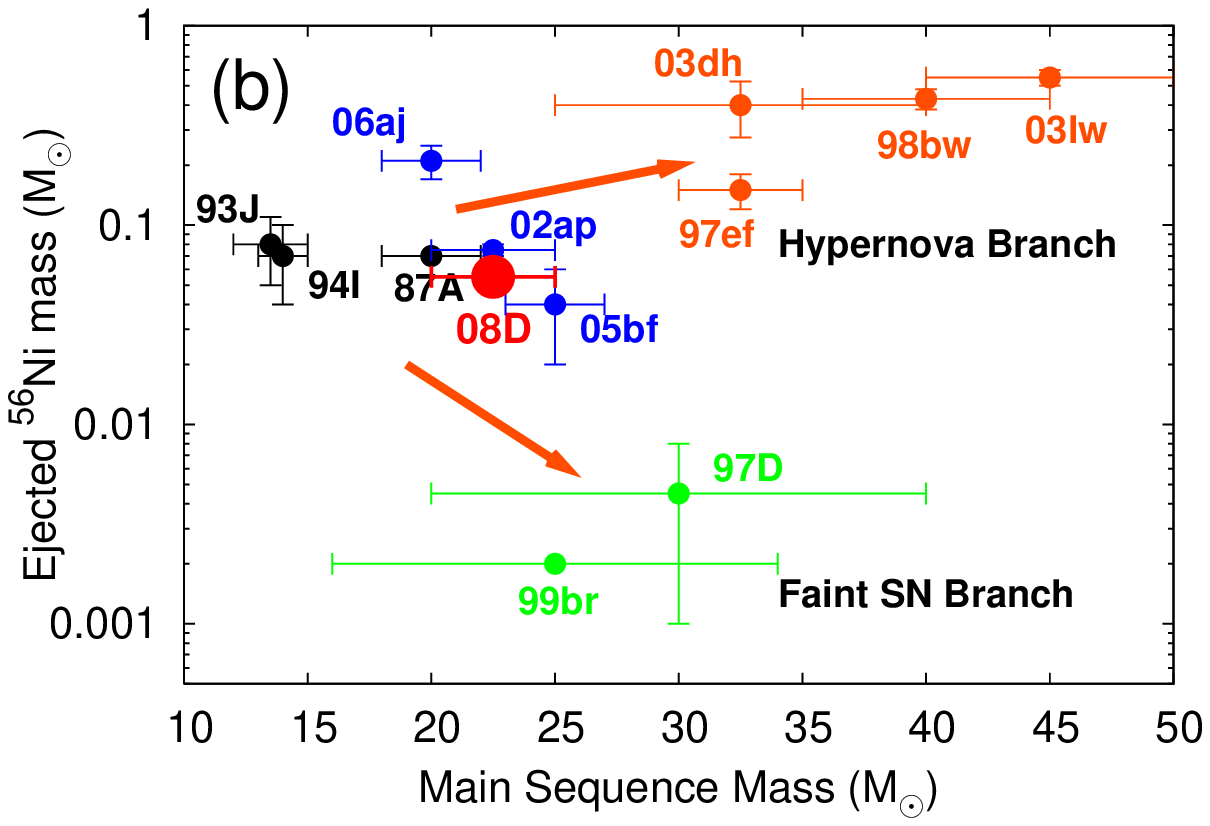}
\caption{
Kinetic energy of the explosion (upper) and the ejected \Nifs\ mass
as a function of the estimated main-sequence mass of the progenitors 
for several core-collapse SNe.
The parameters are listed in Table \ref{tab:param} with references.
The progenitor mass of SNe shown in the figure is estimated 
based on the $\Mms$-$M_{\alpha}$ relation 
by Sugimoto \& Nomoto (1980; used in Nomoto \& Hashimoto 1988) 
as in this paper.
\label{fig:ME}}
\end{center}
\end{figure}

\subsection{Optimal Model for SN 2008D}

For the five He star progenitor models, we calculate hydrodynamics of the 
explosions and explosive nucleosynthesis.
To reproduce the observed LC, 
we obtain the possible set of the mass and kinetic energy of the ejecta:
($\Mej/\Msun$, $\KE/10^{51}$ erg) =
(2.7, 1.1), (4.4, 3.7), (6.2, 8.4), (7.7, 13.0) and (12.4, 26.5)
for HE4, HE6, HE8, HE10 and HE16, respectively.
These five models are tested against the optical spectra.

Model HE4 has many difficulties in reproducing the observed spectra.
At early epochs, the calculated \ion{O}{i} line is too weak
because of the too small oxygen mass in the He-rich layer.
At later epoch, the model spectrum suggests that 
the photospheric layer at $v = 7500$ \kms\ is O-rich, 
which is not consistent with the explosion model that has
He-rich or He-O mixed layers at $v \gsim 3000$ \kms.

Model HE6 can reproduce the observed spectra well.
The evolution of the photospheric velocity calculated with HE6
is in reasonable agreement 
with the velocities derived from the spectral modelling (Fig. \ref{fig:vph}).
The spectral model at $t=32.4$ suggests that the layer at 
$v \sim 7500$ \kms\ is not the He-rich layer,
while the slightly lower $v_{\rm He}$ of HE6 ($6700$ \kms)
implies that the photosphere at this epoch is He-rich.

Model HE8 is reasonably consistent with all the aspects studied in this paper
At all epochs, the optical spectra can be explained 
with a reasonable abundance distribution, 
and the calculated photospheric velocities are consistent with those derived 
from spectrum synthesis.
However, the velocity at the bottom of the He layer ($v_{\rm He}$) in HE8 
is slightly higher than the observed He line velocities.

Model HE10 and HE16 reproduce the early and later spectra reasonably well.
However, these models predict too high photospheric velocity
(Fig. \ref{fig:vph}).
In addition, the velocities at the bottom the He layer
($v_{\rm He}=12,500$ and $17,500$ \kms for HE10 and HE16, respectively), 
are not consistent with the observed line velocity 
($v \sim 9000$ - $10,000$ \kms).

In summary HE4, HE10 and HE16 are not consistent with SN 2008D.
Both HE6 and HE8 have a small inconsistency related to the 
boundary between the He-rich and O-rich layers.
It seems that a model between HE6 and HE8 may be preferable.
However, since there is uncertainty in $v_{\rm He}$ in our model set,
depending on the mass of the He layers,
we include both HE6 and HE8 as possible models.

We conclude that the progenitor star of SN 2008D has a He core mass
$M_{\alpha} = 6-8 \Msun$ prior to the explosion.
This corresponds to a main-sequence
mass of $\Mms =20-25 \Msun$ under the $\Mms$-$M_{\alpha}$ relation
by Sugimoto \& Nomoto (1980; used in Nomoto \& Hashimoto 1988).
We find that SN 2008D is an explosion with $\Mej = 5.3 \pm 1.0 \Msun$ 
and $\KE = 6.0 \pm 2.5 \times 10^{51}$ erg.
The mass of the central remnant is $1.6 - 1.8 \Msun$, which is near 
the boundary mass between the neutron star and the black hole.
Note that the error bars only reflect the uncertainty of the LC
and spectral modelling.
Possible additional uncertainties of the parameters are discussed below.

\begin{deluxetable*}{llllll} 
\tablewidth{0pt}
\tablecaption{Parameters of Supernovae}
\tablehead{
SN (Type) &
$\Mej$ \tablenotemark{a} & 
$\KE$ \tablenotemark{b} &
$M($\Nifs$)$ \tablenotemark{c} &
$M_{\rm MS}$ \tablenotemark{d} & 
Refs.  
}
\startdata
SN 1987A (II pec) & $14.7$          & $1.1\pm 0.3$   & $0.07 $         
& $20 \pm 2$ & 1, 2 \\
SN 1993J (IIb)    & $3.2 \pm 0.3$   & $1.1\pm 0.1$   & $0.08 \pm 0.03$ 
& $13.5 \pm 1.5$  & 3 \\
SN 1994I (Ic)     & $1.05 \pm 0.15$ & $1.0 \pm 0.2$  &  $0.07 \pm 0.03$
& $14 \pm 1$      & 4, 5 \\
SN 1997D (II)     & $\sim 24$        & $\sim 0.4$    & $0.0045 \pm 0.0035$ 
& $30 \pm 10$ & 6 \\
SN 1997ef (Ic)    & $8.6 \pm 1$     & $12.75 \pm 4.75$& $0.15 \pm 0.03$ 
& $32.5 \pm 2.5$ & 7, 8 \\
SN 1998bw (Ic)    & $10.4 \pm 1$    & $40 \pm 10$ \tablenotemark{e}      & $0.43 \pm 0.05$ 
& $40 \pm 5$ & 9, 10 \\
SN 1999br (II)    & $\sim 14$       & $\sim 0.6$       & $0.002$ 
& $25 \pm 9$ & 11 \\
SN 2002ap (Ic)    & $3.25 \pm 0.75$ & $4 \pm 1$        & $0.075 \pm 0.005$ 
& $22.5 \pm 2.5$  & 12 \\
SN 2003dh (Ic)    & $7 \pm 3$       & $35 \pm 15$      & $0.4 \pm 0.125$ 
& $32.5 \pm 7.5$  & 13, 14 \\
SN 2003lw (Ic)    & $\sim 13$       & $55 \pm 5$       & $0.55 \pm 0.05$ 
& $45 \pm 5$      & 15 \\
SN 2005bf (Ib pec)& $6.5 \pm 0.5$   & $1.25 \pm 0.25$  & $0.04 \pm 0.02$ \tablenotemark{f}
& $25 \pm 2$      & 16, 17  \\
SN 2006aj (Ic)    & $1.8 \pm 0.8$   & $2.0 \pm 1.0$    & $0.21 \pm 0.04$ 
& $20 \pm 2$ & 18 \\
SN 2008D (Ib)     & $5.3 \pm 1.0$   & $6.0 \pm 2.5$    & $0.07 \pm 0.005$ 
& $22.5 \pm 2.5$ & this work  \\
\enddata
\tablenotetext{a}{The mass of the SN ejecta ($\Msun$)}
\tablenotetext{b}{The kinetic energy of the SN ejecta ($10^{51}$ erg)}
\tablenotetext{c}{The mass of ejected \Nifs\ ($\Msun$)}
\tablenotetext{d}{Estimated main-sequence mass ($\Msun$)}
\tablenotetext{e}{$\KE = 20 \times 10^{51}$ erg is derived 
from the modelling with a multi-dimensional model 
(in the polar-viewed case, Maeda et al. 2006; Tanaka et al. 2007).}
\tablenotetext{f}{The mass of \Nifs\ is derived from 
the late time observation (Maeda et al. 2007).
The early observations suggest $M$(\Nifs)$=0.3 \Msun$ 
(Tominaga et al. 2005; Folatelli et al. 2006).}
\tablerefs{(1) Shigeyama \& Nomoto (1990), (2) Blinnikov et al. (2000), 
(3) Shigeyama et al. (1994), (4) Iwamoto et al. (1994), 
(5) Sauer et al. (2006), (6) Turatto et al. (1998),
(7) Iwamoto et al. (2000), (8) Mazzali, Iwamoto \& Nomoto (2000),
(9) Nakamura et al. (2001a), (10) Iwamoto et al. (1998),
(11) Zampieri et al. (2003), (12) Mazzali et al. (2002), 
(13) Mazzali et al. (2003), (14) Deng et al. (2005),
(15) Mazzali et al. (2006a), (16) Tominaga et al. (2005), 
(17) Maeda et al. (2007), (18) Mazzali et al. (2006b)
}
\label{tab:param}
\end{deluxetable*}

{\bf Distance and reddening:}
since the distance to the host galaxy and the reddening toward the SN 
include some uncertainties, the ejected \Nifs\ mass could also 
contain $\sim 20$\% uncertainties. 
However, $\Mej$ and $M_{\rm cut}$ are not affected 
because these values are much larger than the ejected \Nifs\ mass.
Thus, the estimated core mass and progenitor mass are not largely affected 
by the uncertainty of the distance and the reddening.

{\bf Asphericity of the explosion:}
possible effects on the estimate of $\Mej$ and $\KE$ 
from asphericity of the ejecta are of interest.
These effects were studied 
for SN 1998bw associated with GRB 980425 
by Maeda et al. (2006) and Tanaka et al. (2007).
They found that the kinetic energy can be smaller
by a factor of $\lsim 2$ in the on-axis case of highly aspherical 
explosion than in the spherical model.
There is little effect in the off-axis case.

Modjaz et al. (2008b) presented the spectrum at $t = 109$ days
and suggested the asphericity of SN 2008D by 
the doubly-peaked emission profile of the \ion{O}{i} line.
Such a profile of the \ion{O}{i} line 
has been interpreted as an off-axis line-of-sight in the axisymmetric
explosion (Maeda et al. 2002; Mazzali et al. 2005; 
Maeda et al. 2008; Modjaz et al. 2008a).
Thus, the estimate by the modelling under spherical symmetry may not 
be largely changed even for the aspherical models.
The quantitative discussion should wait for 
the later spectra, the detailed modelling of the line profile, 
and determination of the degree of asphericity and the line-of-sight.

The effects of asphericity, especially aspherical mass ejection and 
fallback, are also important to determine the relation between the ejected 
\Nifs\ mass and the remnant mass.
The remnant mass in this paper is determined to eject the optimal amount 
of \Nifs\ by one-dimensional hydrodynamic/nucleosynthetic calculations.
However, since the remnant mass could be either larger or smaller depending 
on the asphericity and details of the explosion mechanism,
the estimate by one-dimensional calculations is a reasonable approximation.

{\bf Possible presence of hydrogen:}
Soderberg et al. (2008) identified
the high velocity H$\alpha$ for the absorption line at 6150 \AA.
If the mass of the H layer is not negligible, it might affect
the core mass, which we estimate by assuming non-existence of H, 
\ie a bare He core.
However, we find a large mass of the H layer is inconsistent with 
the spectrum at $t = 4.6$ days (Appendix \ref{app:H}).
The mass of the H layer is smaller than $5 \times 10^{-4} \Msun$,
and thus, there is no effect on the parameters.

{\bf Evolutionary models:}
the estimate of the main-sequence mass uses 
the approximate $\Mms$-$M_{\alpha}$ relation by Sugimoto \& Nomoto 
(1980; Eq. 4.1), which is used in Nomoto \& Hashimoto (1988).
The $\Mms$-$M_{\alpha}$ relation of several evolutionary models
are shown in Figure \ref{fig:ms_core}.
The systematic differences in this relation 
for $\Mms \lsim 30 \Msun$ ($M_{\alpha} \lsim 10 \Msun$)
may stem from the differences in the treatment of convection, mass loss,
rotation, and binary effects (\S \ref{sec:progenitor}).
Thus, we should keep in mind that 
the main-sequence mass is subject to 
systematic uncertainties of $3-5 \Msun$ (Fig. \ref{fig:ms_core}).
Note that our estimate of the He core mass depends only on 
the estimates of $\Mej$ and $\KE$
from hydrodynamic/nucleosynthetic calculations, and thus, 
our determination of
the He core mass is not affected by the variety of the evolutionary models.

\begin{figure}
\begin{center}
\includegraphics[scale=1.1]{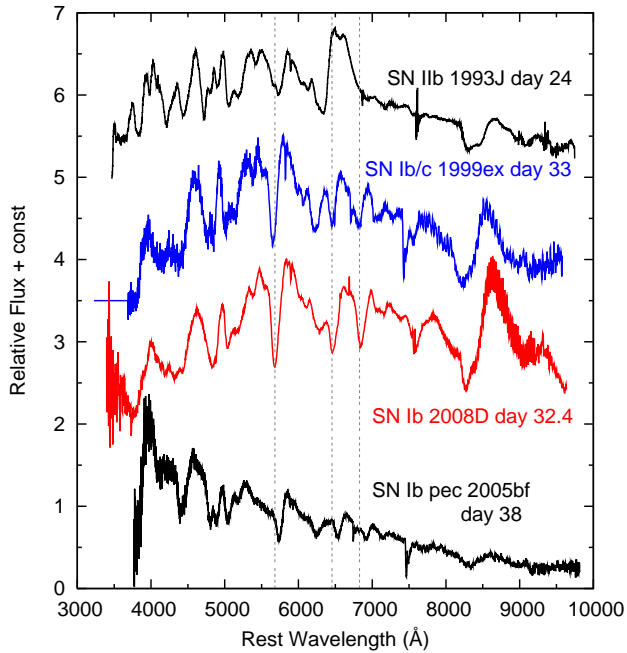}
\caption{
Spectral comparison among SNe 1993J (IIb; Barbon et al. 1995), 
1999ex (Ib/c; Hamuy et al. 2002), 2008D (Ib) 
and 2005bf (peculiar Ib; Anupama et al. 2005).
The epoch for SNe 1993J, 1999ex, and 2005bf 
is given in the estimated days from the explosion.
Dashed, vertical lines show the position of the He lines 
(\ion{He}{i} 5876, 6678, 7065) blueshifted with $v=10,000$ \kms.
The spectra except for SN 2008D are taken from the SUSPECT database.
\label{fig:comp}}
\end{center}
\end{figure}

\subsection{Comparison with Previous Works}
\label{sec:comp}

Soderberg et al. (2008) have estimated the parameters of the ejecta as
$\Mej = 3-5 \Msun$ and $\KE = 2-4 \times 10^{51}$ ergs,
which are smaller than those derived in this paper.
The difference seems to stem from their assumptions of 
the homogeneous sphere and time-independent opacity.
These assumptions lead an almost time-independent photospheric velocity,
which is not the case in SNe.
Especially for SN 2008D, the very early spectra show the broad-line features, 
and the photospheric velocity at $t \sim 5$ days after the X-ray transient
is almost twice as high as the velocity around maximum.

Adopting the cooling envelope model by Waxman et al. (2007) to the 
blackbody temperature and the radius at $t \lsim 4$ days,
Soderberg et al. (2008) estimated a progenitor radius to be
$R_* \sim 1 \Rsun$ with $E(B-V)=0.61$ mag, 
$\Mej=5 \Msun$ and $\KE=2 \times 10^{51}$ erg.
Modjaz et al. (2008b) also derived a similar value, 
$R_* = 1.1 \pm 0.46 \Rsun$ with $E(B-V)=0.6$ mag 
and the same $\Mej$ and $\KE$ with Soderberg et al. (2008).
If $\Mej$ and $\KE$ derived in this paper are adopted, 
the estimated radius is $\sim 80 \%$ of their estimate.
This is marginally consistent with the radius of model HE8 
while it is smaller than HE6.
In this sense, model HE8 seems to be more self-consistent.
It must be noted, however, that Chevalier \& Fransson (2008) derived 
a larger radius, $R_* \sim 9 \Rsun$ by using the model by Chevalier (1992,
and using the blackbody temperature and the radius presented by
Soderberg et al. 2008).

Mazzali et al. (2008) estimated the ejecta parameters 
by modelling the bolometric LC and optical spectra.
Their largest assumption is that a central remnant as massive
as $\sim 3 \Msun$ is implicitly assumed,
which leads a massive He core mass ($\sim 10 \Msun$), and thus,
a massive progenitor mass ($\sim 30 \Msun$).
Our hydrodynamic/nucleosynthetic calculations show that a smaller central 
remnant is preferred ($\sim 1.6-1.8 \Msun$, Table \ref{tab:models}).

\subsection{SN 2008D in the Context of Type Ib/c Supernovae}
\label{sec:Ibc}

Figure \ref{fig:ME} shows the kinetic energy of the ejecta and the ejected
\Nifs\ mass as a function of the estimated main-sequence mass for several
core-collapse SNe (see, \eg Nomoto et al. 2007). 
The parameters shown in Figure \ref{fig:ME} are also listed in 
Table \ref{tab:param}.
SN 2008D is shown by a red circle in Figure \ref{fig:ME}.
The ejecta parameters for other SNe shown in Figure \ref{fig:ME} 
and Table \ref{tab:param} are derived from one-dimensional modelling 
as in this paper.
Although there is a systematic uncertainty in the progenitor mass 
(Fig. \ref{fig:ms_core}), 
the progenitor mass of SNe shown in Figure \ref{fig:ME} is estimated 
based on the $\Mms$-$M_{\alpha}$ relation 
by Sugimoto \& Nomoto (1980; used in Nomoto \& Hashimoto 1988) 
as in this paper.
Thus, the relative position of SNe in the plots are robust.

The main-sequence mass of the progenitor of SN 2008D is estimated to
be between normal SNe and GRB-SNe (or hypernovae).
The kinetic energy of SN 2008D is also intermediate.
Thus, SN 2008D is located between the normal SNe and the ``hypernovae branch''
in the $\KE\ - \Mms$ diagram (upper panel of Fig. \ref{fig:ME}).
The ejected \Nifs\ mass in SN 2008D ($\sim 0.07 \Msun$)
is similar to the \Nifs\ masses ejected by normal SNe and much smaller than
those in GRB-SNe.

Figure \ref{fig:comp} compares the spectra of
SNe 1993J (IIb, Barbon et al. 1995), 1999ex (Ib/c, Hamuy et al. 2002), 
2005bf (Ib, Anupama et al. 2005, Tominaga et al. 2005, Folatelli et al. 2006), 
and 2008D (Ib).
The epoch for SNe 1993J, 1999ex, and 2005bf 
is given in the estimated days from the explosion.
The explosion epoch is uncertain up to $\sim 15$ days in SN 2005bf
(Folatelli et al. 2006)
while it is well constrained in SNe 1993J and 1999ex ($\lsim 2$ days,
Wheeler et al. 1993; Hamuy et al. 2002).

The spectra of SN 2008D and SN 1999ex are very similar (Valenti et al. 2008b),
while SN 2005bf has a lower He velocities.
Although the epoch of SN 2005bf is uncertain, 
the He line velocities in SN 2005bf is always lower than 8000 \kms\
(Tominaga et al. 2005).
The He lines in SN 1993J are very weak at this epoch.
The Fe features at 4500-5000\AA\ are similar in these four SNe,
but those in SN 2005bf are narrower.

Malesani et al. (2009) suggested that 
the bolometric LCs of SNe 1999ex and 2008D are similar.
If it is the case (although some discrepancy is shown by Modjaz et al. 2008b),
the similarity in both the LC and the spectra suggests that SN 1999ex is 
located close to SN 2008D in the $\KE\ - \Mms$ and $\Mni - \Mms$ diagrams.

Comparison with other Type Ib SNe shown in Figure \ref{fig:ME} is 
possible only for SN 2005bf although SN 2005bf is 
a very peculiar SN that shows a double peak LC with 
a very steep decline after the maximum, and increasing He line velocities 
(Anupama et al. 2005; Tominaga et al. 2005; Folatelli et al. 2006; 
Maeda et al. 2007).
The LC of SN 2005bf is broader than that of SN 2008D,
while the expansion velocity of SN 2005bf is lower than that of SN 2008D.
These facts suggest 
that SN 2005bf is the explosion with lower $\KE/ \Mej$ ratio
(Table \ref{tab:param}).

Malesani et al. (2009) also pointed the similarity of 
the LCs of SNe 1993J and 2008D. 
But the expansion velocity is higher in SN 2008D 
(see, \eg Barbon et al. 1995; Prabhu et al. 1995).
Thus, both the mass and the kinetic energy of 
the ejecta are expected to be smaller in SN 1993J.
In fact, SN 1993J is explained by the explosion of a $4 \Msun$ He core
with a small mass H-rich envelope
(Nomoto et al. 1993; Shigeyama et al. 1994; Woosley et al. 1994).

\section{Conclusions}
\label{sec:conclusions}

We presented a theoretical model for SN 2008D associated 
with the luminous X-ray transient 080109.
Based on the progenitor models,
hydrodynamics and explosive nucleosynthesis are calculated.
Using the explosion models, radiative transfer calculations are performed.
These models are tested against the bolometric LC and optical spectra.
This is the first detailed model calculation for Type Ib SN
that is discovered shortly after the explosion.

We find that SN 2008D is a more energetic explosion than
normal core-collapse SNe.
We estimate that the ejecta mass is $\Mej = 5.3 \pm 1.0 \Msun$
and the total kinetic energy of $\KE = 6.0 \pm 2.5 \times 10^{51}$ erg.
The ejected \Nifs\ mass is $\sim 0.07 \Msun$.
To eject the optimal amount of \Nifs,
the mass of the central remnant is estimated to be $1.6 - 1.8 \Msun$.
The error bars include only the uncertainty of the LC and spectral modelling.

Summing up the above masses, it is concluded that 
the progenitor star of SN 2008D has a $6 - 8 \Msun$ He core
prior to the explosion.
There is essentially no H envelope with the upper limit of 
$5 \times 10^{-4} \Msun$.
Thus, the corresponding main-sequence mass
of the progenitor is $\Mms =20-25 \Msun$ under 
the $\Mms$-$M_{\alpha}$ relation by Sugimoto \& Nomoto 
(1980, used in Nomoto \& Hashimoto 1988).
We note that there exist additional systematic uncertainties
in this relation due to convection, mass loss, rotation, and binary effects.
Our estimates of these masses and energy suggest that SN 2008D is near 
the border between neutron star-forming and black hole-forming SNe, and has 
properties intermediate between those of normal SNe and 
hypernovae associated with gamma-ray bursts.

\acknowledgments
M.T. and N.T. are supported by the JSPS 
(Japan Society for the Promotion of Science) 
Research Fellowship for Young Scientists.
We have utilized the SUSPECT database.
We would like to thank the contributors of the spectra used in the paper.
This research has been supported in part by World Premier International
Research Center Initiative (WPI Initiative), MEXT, Japan, and by
the Grant-in-Aid for Scientific Research of the JSPS 
(10041110, 10304014, 11740120, 12640233, 14047206, 14253001, 14540223, 
16740106, 18104003, 18540231, 20540226) and MEXT
(19047004, 20040004, 20041005, 07CE2002).

\appendix

\section{A. Construction of Bolometric Light Curve}
\label{app:LC}

The bolometric LC shown in this paper was constructed 
by using optical data taken by the MAGNUM telescope 
(Yoshii 2002; Yoshii, Kobayashi \& Minezaki 2003), 
the Himalayan Chandra Telescope, 
and {\it Swift} UVOT (U-band, Soderberg et al. 2008),
and also NIR data taken by the MAGNUM telescope.

The bolometric luminosity was derived integrating 
the flux from $U$ (with the edge of $9.68 \times 10^{14}$ Hz)
to $K$ ($1.00 \times 10^{14}$ Hz) band.
The photometric points are interpolated by the third order natural spline.
If the data point of a certain band is not available, 
we use linear interpolation of the magnitude.

The derived bolometric LC can be compared with that by 
Soderberg et al. (2008), Malesani et al. (2009),
Modjaz et al. (2008b, $U$-$K_S$ integration) 
and Mazzali et al. (2008).
Although the scatter up to 0.4 mag is found 
among the LCs by direct comparison, 
it is caused mainly by the difference in the assumed distance and 
reddening.

If the same distance and reddening are used 
(here we approximately correct the difference 
in the bolometric magnitude $\Delta M_{\rm bol}$ 
caused by the difference in the assumed reddening by
$\Delta M_{\rm bol} = R_{V} \Delta E(B-V)$, where $ R_{V}$=3.1 and 
$\Delta E(B-V)$ is the difference in the assumed color excess), 
the LCs around/after the maximum is consistent 
among those by this paper, Soderberg et al. (2008),
Modjaz et al. (2008b) and Mazzali et al. (2008) within 0.1 mag,
while the LC by Malesani et al. (2009) is fainter by 0.2-0.5 mag.
For the pre-maximum epochs, the LCs by these papers are consistent 
within 0.2 mag 
except that the magnitude at $t=4$ days by this paper 
(shown by the arrow in the left panel of Fig. \ref{fig:LC})
is brighter than other ones by 0.25 mag.

Since the scatter in the maximum luminosity among the papers 
is up to 0.2 mag, it causes the uncertainty of the ejected \Nifs\ mass
up to $\sim 20 \%$.
However, this uncertainty does not affect 
our determination of the ejecta mass 
because the change in the \Nifs\ mass (and mass cut)
is negligible compared to the ejecta mass (\S \ref{sec:discussion}).
In addition, the time scale of the bolometric LC around the maximum is 
reasonably consistent among the papers,
the kinetic energy of the ejecta is also not affected.

\section{B. Nonexistence of the Hydrogen Layers}
\label{app:H}

\begin{figure*}
\begin{center}
\begin{tabular}{cc}
\includegraphics[scale=1.0]{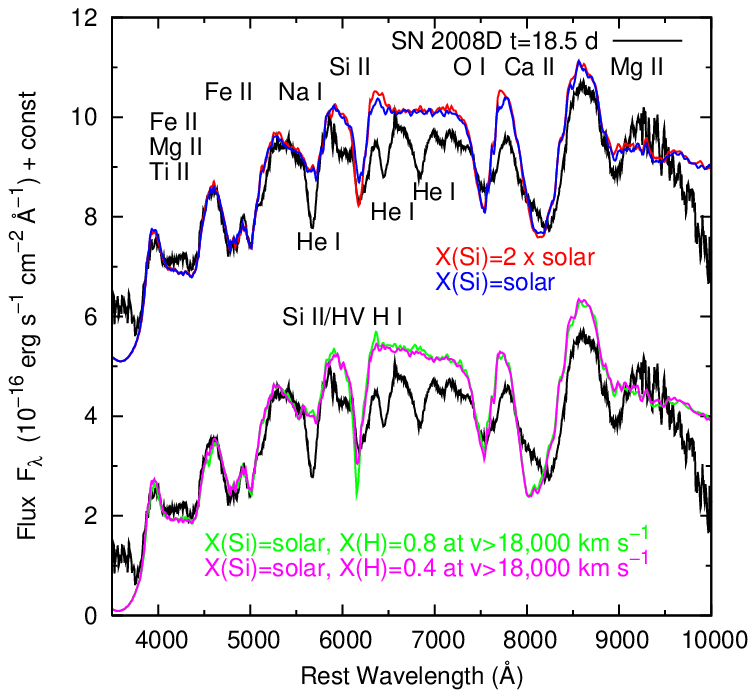}&
\includegraphics[scale=1.0]{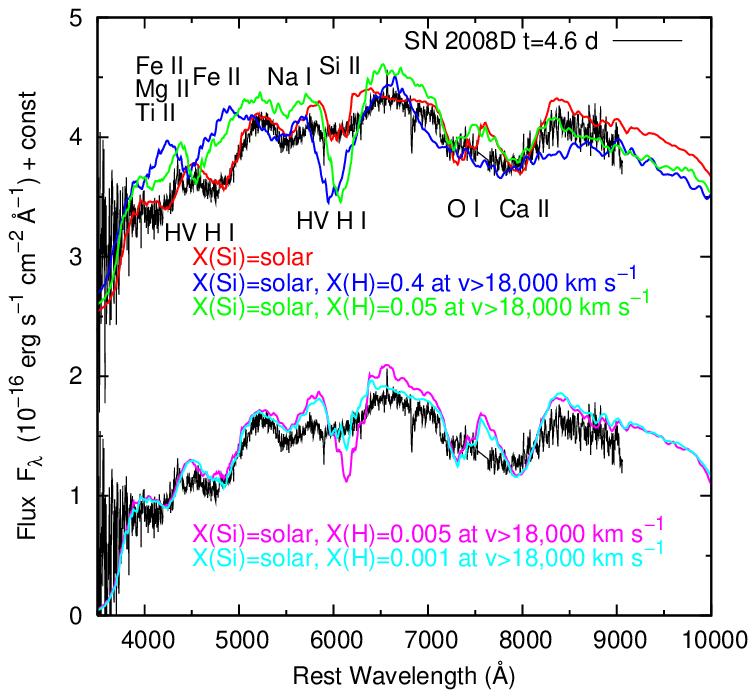}
\end{tabular}
\caption{
{\it Left}: Optical spectrum at $t=18.5$ days (Mazzali et al. 2008)
compared with the synthetic spectra.
The red and blue lines show the synthetic spectra with the models where
the Si mass fraction is twice as large as the solar abundance (best fit) and 
the same as the solar abundance, respectively.
The element abundances are assumed to be homogeneous in these models.
The green and magenta lines show the synthetic spectra with 
$X$(H)=0.8 and 0.4 at $v > 18000$ \kms, respectively.
The corresponding mass of H is 0.4 and 0.2 $\Msun$, respectively.
The homogeneous, solar abundance of Si is assumed in these models.
{\it Right}: The optical spectrum at $t=4.6$ days (Mazzali et al. 2008)
compared with the synthetic spectra.
The red line shows the model with the solar abundance of Si, which gives
a good fit to the observed spectrum.
The green, blue, magenta and cyan lines show the models with 
$X$(H) = $0.4, 0.05, 0.005$ and $0.001$ at  $v > 18000$ \kms, respectively.
The corresponding mass of H is 0.4, 0.025, 0.0025 and 0.0005 $\Msun$, 
respectively.
\label{fig:H}
}
\end{center}
\end{figure*}

Soderberg et al. (2008) identified 
the high velocity (HV) H$\alpha$ line for the absorption feature 
around 6150 \AA\ in the spectra around maximum.
It is blended with the strong \ion{Si}{ii} line, and 
discrimination is not easy (\eg Branch et al. 2006; Elmhamdi et al. 2006).
The presence of H is important to specify the 
properties of the progenitor star just prior to the explosion.
In addition, if the H layer is present, 
the estimate of $\Mej$ and $\KE$ may be affected 
since we have used bare He stars for the LC and spectral modelling.

First, we test the presence of H in the spectrum around maximum
using model HE8.
The left panel of Figure \ref{fig:H} shows the comparison between 
the observed spectrum at $t=18.5$ days (Mazzali et al. 2008) 
and synthetic spectra.
The photospheric velocity at this epoch is 9000 \kms\ (Fig. \ref{fig:vph}).
If the absorption at 6150 \AA\ is \ion{Si}{ii} $\lambda$6355, 
the Doppler velocity of the absorption at 6150 \AA\ is $9300$ \kms, 
which is well consistent with the photospheric velocity.
The red line shows the best fit model that includes Si twice as large 
as the solar abundance.
The absorption is slightly shallower in the model 
with solar abundance Si (blue).
Since the abundance twice as large as the solar abundance is reasonable 
for the middle layers of the ejecta, 
the HV \ion{H}{i} is not necessarily required.

However, this does not exclude the possibility of the presence of H 
at the outer layers.
If the absorption at 6150 \AA\ is H$\alpha$, 
the Doppler velocity is $18500$ \kms.
To test the presence of the H at such high velocity layers,
we calculate model spectra by replacing He at $v>18000$ \kms\ with H.
The green and magenta lines show the models with $X$(H)=0.8 and 0.4 
at $v>18000$ \kms, respectively.
The corresponding mass of H is 0.4 and 0.2 $\Msun$, respectively.
The models also include the solar abundance of Si at at $v > 9000$ \kms.
While the model with $X$(H)=0.8 gives a too strong absorption, 
the model with $X$(H)=0.4 agree with the observed spectrum.
Thus, the presence of 0.2 $\Msun$ of H cannot be denied from 
the spectrum around the maximum.

Next, we perform the similar tests using the very early spectrum.
The right panel of Figure \ref{fig:H} shows the comparison between 
the observed spectrum at $t=4.6$ days (Mazzali et al. 2008) 
and the synthetic spectra.
The red line shows the best fit model, which include the solar abundance of Si.
The blue line shows the model that have the solar abundance of Si and 
$X$(H)=0.4 at $v>18000$ \kms.
Although this model gives a reasonable fit to the 
spectrum at $t$=18.5 days (left panel of Fig. \ref{fig:H}),
it shows too strong H$\alpha$ and H$\beta$ at $t=4.6$ days
(the lack of H$\beta$ has been pointed out by Malessani et al. 2009).
We get the stronger line at earlier epochs because 
the density at the high velocity layers ($v=18000$ \kms) becomes
lower with time, and the line forming there is more effective at
earlier epochs.

The green, magenta and cyan lines show the models with smaller mass fraction
of H, $X$(H)=0.05, 0.005, and 0.001, respectively.
The corresponding mass of H is 0.025, 0.0025 and 0.0005 $\Msun$.
The models with $X$(H) = 0.05 and 0.005 (green and magenta) 
still shows too strong \ion{H}{i} lines.
With $X$(H) = 0.001, the H$\alpha$ line has little effect 
on the absorption at 6150 \AA\ although the model spectrum still has 
a sharp absorption of the HV H$\alpha$.

If we use model HE6, the mass at the outer layers is smaller.
Thus, we conclude that the mass of H is smaller than $5 \times 10^{-4} \Msun$.

\end{document}